\documentclass[oldversion]{aa}  
\usepackage{graphicx}
\usepackage{txfonts}
\usepackage{aalongtable}
\begin{document}

   \title{Non-thermal radio emission from O-type stars}
   \subtitle{II. HD 167971}

   \author{R. Blomme \inst{1}
           \and M. De Becker \inst{2}
           \and M.C. Runacres \inst{3}
           \and S. Van Loo \inst{4}
           \and D. Y. A. Setia Gunawan \inst{5}
          }

   \offprints{R. Blomme,\\ \email{Ronny.Blomme@oma.be}}

   \institute{Royal Observatory of Belgium,
              Ringlaan 3, B-1180 Brussel, Belgium
            \and
              Institut d'Astrophysique, Universit\'e de Li\`ege,
              All\'ee du 6 Ao\^ut, 17, B\^at B5c,
              B-4000 Li\`ege (Sart-Tilman), Belgium
            \and
              Vrije Universiteit Brussel,
              Pleinlaan 2, B-1050 Brussel, Belgium
            \and
              School of Physics and Astronomy,
              The University of Leeds, Woodhouse Lane, Leeds LS2 9JT, UK
            \and
              Australia Telescope National Facility,
              PO Box 76, Epping, NSW 2121, Australia
             }

   \date{Received date / accepted date}

   \abstract{
HD~167971 is a triple system consisting of a 3.3-day eclipsing binary
(O5--8 V + O5--8 V) and an O8 supergiant. It is also a well known non-thermal
radio emitter. 
We observed the radio emission of HD~167971
with the Very Large Array (VLA) and the Australia
Telescope Compact Array (ATCA). By combining these data with VLA archive
observations we constructed a radio lightcurve covering a 20-yr
time-range.
We searched for, but failed to find, the 3.3-day spectroscopic
period of the binary in the radio data.
  This could be due to the absence of intrinsic synchrotron radiation
  at the colliding-wind region between the two components of the eclipsing 
  binary,
  or due to the large 
  amount of free-free absorption that blocks the synchrotron 
  radiation.
We are able to explain many of the observed characteristics of the radio data
     if the non-thermal emission is produced in a colliding-wind region
     between the supergiant and the combined winds of the binary. 
     Furthermore, if
the system is gravitationally bound, the 
orbital motion occurs over a period
of $\sim$\,20 years, or longer, as suggested by the long-term variability
in the radio data. 
We argue that the variability is due to the free-free
absorption that changes with orbital phase or may also in part be due to
changes in separation, should the orbit be eccentric.

   \keywords{stars: individual: HD 167971 --
             stars: early-type --
             stars: mass-loss --
             stars: winds, outflows --
             radio continuum: stars --
             radiation mechanisms: non-thermal
               }
}

\titlerunning{Non-thermal radio emission from O-type stars. II}
\authorrunning{R. Blomme et al.}

   \maketitle
%

\section{Introduction}

Radio emission from most hot stars is due to thermal free-free emission
by material in their stellar winds. A significant fraction of these stars,
however, also show evidence of non-thermal emission.
The radio fluxes of non-thermal sources are
characterized by a zero or negative spectral
index\footnote{The radio spectral index $\alpha$ describes the power-law
      behaviour of the flux:
      $F_\nu \propto \nu^\alpha \propto \lambda^{-\alpha}$.
              }
 and a high brightness temperature.

For Wolf-Rayet stars, Dougherty \& Williams (\cite{Dougherty+Williams00})
showed that non-thermal emission is strongly correlated with binarity.
In a binary, the stellar winds of both components collide.
  On either side of the surface where the wind pressures balance,
  shocks are formed that define the extent of the colliding-wind region
  (Eichler \& Usov~\cite{Eichler+Usov93}). Around these shocks,
relativistic electrons are accelerated
by the first-order Fermi mechanism (Bell~\cite{Bell78}).
These relativistic electrons then
spiral in a magnetic field and thereby
emit synchrotron radiation, which we detect as non-thermal radiation.

For O stars, the link between non-thermal emission and binarity was,
until recently, less clear. 
There are a number of spectroscopically single O stars that are
non-thermal emitters.
In a single star, relativistic electrons could be accelerated
in shocks that are due to the instability of the radiative
driving mechanism (e.g. Owocki et al.~\cite{Owocki+al88}).
However, Van Loo et al.~(\cite{VanLoo+al06}) showed that this embedded
shock model cannot explain the observed spectral index. 
They conclude that, 
     just as for the Wolf-Rayet stars, all 
     O-star non-thermal emitters
should be binaries. 
Non-thermal emitters that are apparently single
must therefore be binaries with an unfavourable inclination angle
or a long-period orbit, making spectroscopic detection of their
binary nature difficult.
An example of such an object is \object{HD~168112}, which we
studied in a previous paper 
(Blomme et al.~\cite{Blomme+al05}, hereafter Paper~I).
The periodic nature of the non-thermal radio fluxes suggests that
this spectroscopically single star is in fact a binary,
with a period estimated to be 1.4~yr.

Problems remain, however, in understanding the observed radio emission
in these colliding-wind binaries. \object{Cyg~OB2 No.~5}, for example, is
a 6.6-day period binary where a colliding-wind region is clearly seen
(in high-resolution VLA observations) between the binary and a third star.
The radio flux of the binary
alternates between a ``high" flux state
and a ``low" flux state and it does so on a time-scale of $\sim$\,7~years,
rather than the 6.6-day orbital period
(Miralles et al.~\cite{Miralles+al94}, Contreras et al.~\cite{Contreras+al97}).
To improve our understanding of the radio emission in binaries,
a more detailed study of this type of object is therefore warranted.

In this paper we study the non-thermal radio emitter \object{HD~167971}
(RA = $18^{\rm h}18^{\rm m}05{\fs}895$, 
Dec = $-12${\degr}14{\arcmin}33{\farcs}31, J2000).
Leitherer et al.~(\cite{Leitherer+al87})
found that this system consists of a 3.3213-day period eclipsing binary
and a third companion. 
The binary components
are very similar to one another in mass and temperature, but the 
spectral types are not well determined: they can only be constrained 
to the range O5--O8. 
 From the absolute magnitudes, Leitherer et al. tentatively estimate
  both components to be main sequence. 
  From a more detailed modelling of the lightcurve,
  Davidge \& Forbes~(\cite{Davidge+Forbes88}) propose a giant or supergiant
  classification.
The spectral lines
also show the third component to be 
a more luminous star, probably an
O8 supergiant, and it is not known if it is gravitationally bound
to the system or is just a line-of-sight object. 
 No high spatial resolution observations are available, so the
  angular separation between the binary and the third component 
  is not known.
Further orbital parameters of the binary
are not well determined: the eccentricity is probably
$e \approx 0$.
No detailed coverage of radial velocities as a function of phase in
the binary orbit is available, but spectra taken near quadrature indicate
a velocity
amplitude of $\sim$\,300~km s$^{-1}$. 
No significant velocity variations
in the spectral lines of the third component have been detected.
The distance to HD~167971 can be derived from its membership of the
open cluster \object{NGC~6604}, which is part of the 
\object{Ser~OB2} association. From
the distance modulus listed by
Humphreys (\cite{Humphreys78}) a distance of 2 kpc is found.

The purpose of this paper is to study the colliding-wind regions
in the HD~167971 system, by combining new radio observations with
archive data.
We construct a simple model that gives a plausible explanation
for the important features seen in the radio fluxes
and that is also compatible with the X-ray and VLBI
(Very Long Baseline Interferometry) observations.
Detailed models, such as those developed
by Dougherty et al.~(\cite{Dougherty+al03}) and
Pittard et al.~(\cite{Pittard+al06}) are not used here.
Many of the orbital and stellar parameters are not known
(especially for the interaction between the binary and the supergiant),
making the parameter space too large for sophisticated modelling.

The rest of this paper is structured as follows. 
In Sect.~\ref{section data} we present the radio observations,
with the details of the data reduction split off to
Appendix~\ref{appendix data reduction}.
We discuss the variability detected in the radio in 
Sect.~\ref{section variability}. The observations are interpreted in
Sect.~\ref{section interpretation} and the conclusions are drawn
in Sect.~\ref{section conclusions}.

\section{Data}
\label{section data}

We observed HD~167971 with the NRAO Very Large
Array\footnote{The National Radio Astronomy
             Observatory is a facility of the National Science Foundation
             operated under cooperative agreement by Associated Universities,
             Inc.}
(VLA) on 2002 March 24 (at 3.6, 6 and 20~cm) and 2002 September 11
(at 3.6, 6, 18 and 20~cm). All observations were centred near HD~167971,
except the 2002 September 11 observations at 18 and 20~cm, which were
centred near HD~168112. The field covered at these wavelengths is large
enough to contain HD~167971 as well. All observations consist of a single
run on the target, preceded and followed by a phase calibrator. The details
of the reduction are given in Appendix~\ref{appendix data reduction}.

In Paper~I we described the Australia Telescope Compact
Array\footnote{The Australia Telescope Compact Array is part of the
          Australia Telescope which is funded by the Commonwealth of
          Australia for operation as a National Facility managed by
          CSIRO.}
(ATCA) observations which were collected on 2001 October 11, centred on
HD~168112. The 13 and 20~cm observations have a field that is large enough
to contain HD~167971. Table~\ref{table radio data}
gives more information about the reduction of this dataset and the fluxes we
determined.

\begin{figure*}
\resizebox{\hsize}{!}{\includegraphics{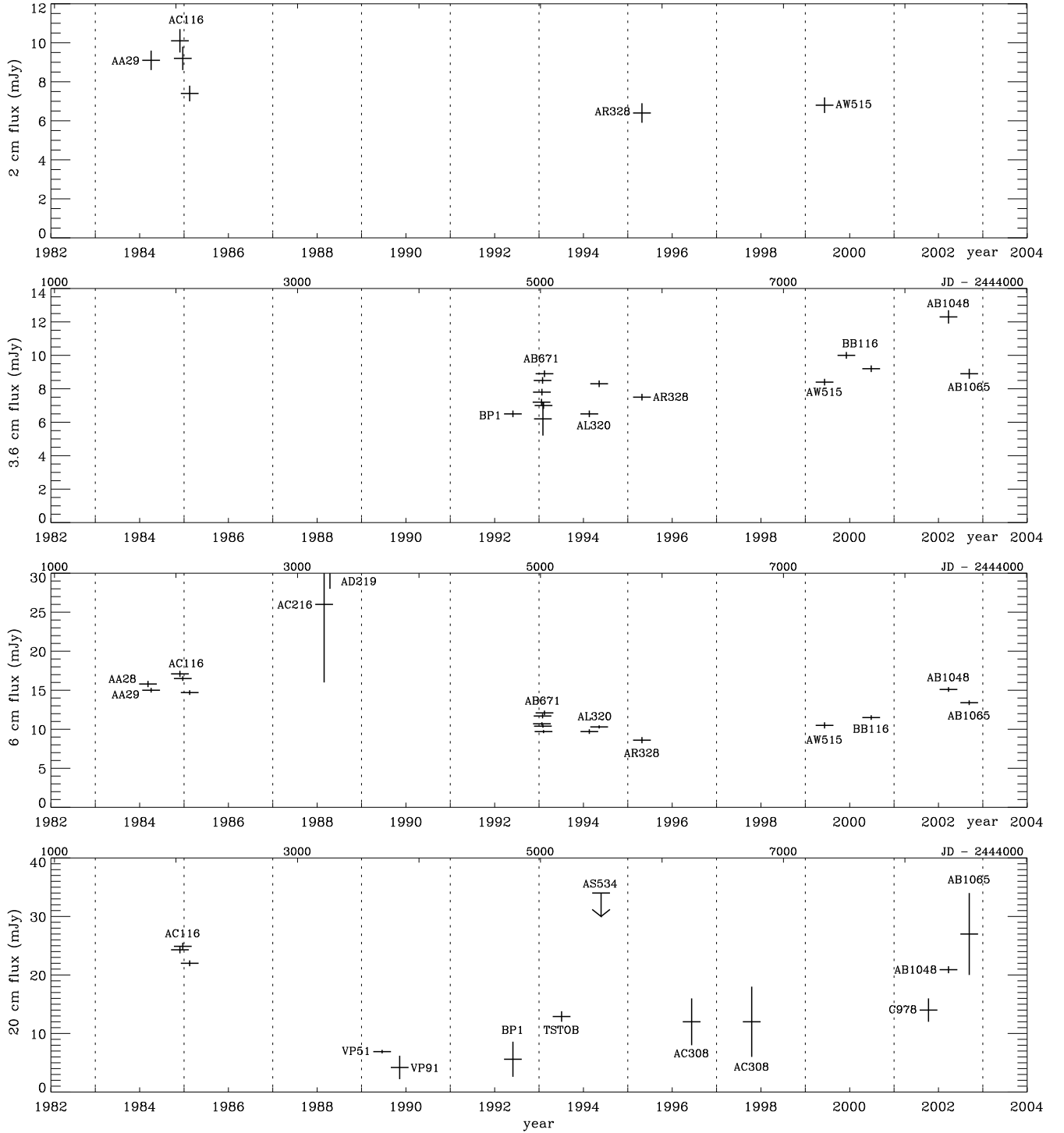}}
\caption{The 2, 3.6, 6 and 20~cm radio fluxes of HD~167971,
plotted as a function of time. 
  Not all 20~cm upper limits are included in this figure, only those which
  are lower than 40~mJy. Fluxes at 0.7, 1.3, 13 and 90~cm are not plotted, 
  because only one or two data points are available. Note that there are 
  differences of a few percent in wavelength for observations we classed in 
  the same band (these differences are up to 20~\% in the 20~cm band).
Observations are
labelled by their programme name.
}
\label{figure fluxes}
\end{figure*}

We supplemented our own observations with archive data.
The VLA archive contains a number of observations that are centred on,
or close to, HD~167971. (The ATCA archive was also searched, but no additional
data were found.) 
Many of these observations have not been published previously.
To avoid introducing systematic effects in the data reduction,
we re-reduced all observations consistently.
Details of the reduction and the measured fluxes are given
in Appendix~\ref{appendix data reduction}.
 The 2, 3.6, 6 and 20~cm fluxes are plotted in Fig.~\ref{figure fluxes}.

Fig.~\ref{figure fluxes} confirms the non-thermal nature of the radio
emission from HD~167971. The fluxes 
increase with larger wavelength (with some exceptions at 20~cm) and 
 the spectral index is therefore negative.
The fluxes are also clearly variable and very high compared
to what would be expected from thermal free-free emission. 
  The stellar wind of the O8 supergiant
  dominates the free-free emission, with a thermal flux of only 0.1~mJy
  at 6~cm based on typical wind parameters listed in 
  Table~\ref{table typical parameters}a.

\section{Variability}
\label{section variability}

\subsection{Possible instrumental effects}
\label{section possible instrumental effects}

 A number of problems arose during the data reduction (see 
  Appendix~\ref{appendix data reduction}). One of those was that many 
  observations have a phase calibrator at a rather large angular distance 
  from the target (9-11\degr), which forced us to apply selfcalibration
  to almost all observations. It is therefore important to check
that the variability seen in Fig.~\ref{figure fluxes}
is not instrumental. 
 First of all, the HD~167971 flux should not be correlated
  with the flux derived for the phase calibrator.
  The phase calibrator fluxes were determined
  from the AIPS task {\sc getjy} (except for the ATCA data, where we measured
  them on an image) and are listed in Table~\ref{table radio data}.
  We checked those calibrators
  that were used in three or more observations
  and found
  that the phase calibrator and the HD~167971 flux are not significantly
  correlated.

  Secondly, any flux changes of other targets
  on the same image should not be correlated with
  the flux changes of HD~167971.
This test could only be applied 
at 6~cm (one other target) and 20~cm (three other targets).
  Again, when we tested this, no significant correlation was found.
  Furthermore, the range in variability in HD~167971
  is substantially larger than that seen in the other targets.

The VLA instrument can be used in a number of configurations that
correspond to different spatial resolutions.
We therefore also need to check that the higher resolution observations
do not partly resolve the target and therefore result in a lower flux.
     From VLBI observations, 
     Phillips \& Titus~(\cite{Phillips+Titus90}) found
     that the linear size of the non-thermal region is at most 16 
     milli-arcsec (mas) at the time of their observation,
which is well within the beamsize of the VLA
($\sim$\,0{\farcs}5 at 6~cm, in the highest resolution configuration). 
Furthermore, at 6~cm, the highest-resolution A configuration gives both the 
highest
flux value and a very low one (see Table~\ref{table radio data}). At 20~cm, the
A configuration gives the highest value, while the low values were obtained
rather with the lower-resolution C configuration.
The observed variability
can therefore not be ascribed to our partly resolving HD~167971.

\subsection{Long-term variability}
\label{section long-term variability}

It is quite obvious from Fig.~\ref{figure fluxes} that the fluxes show
long-term variability. This is most clearly seen at 6 and 20~cm.
The 6~cm fluxes vary between $\sim$\,8 and
$\sim$\,18~mJy 
 (or even higher, if we include the less reliable AC216 and AD219 
  observations) and
the 20~cm fluxes vary between $\sim$\,4 and $\sim$\,25~mJy
(over a time-scale of 5--10 years).

The 6~cm dataset is the most complete, in the sense that it covers the
$\sim$\,20-yr time-range reasonably well.
Judging by eye, a sine function seems quite appropriate to fit 
the long-term trend in these
data (although this is based in part on 
the 1988 AC216 and AD219 observations, which have large error bars).
Fitting such a sine function, we found a $\sim$\,20-yr period,
with a maximum around 1986.2.
Of course, a period that is so similar to the time-range covered by the
observations is not really convincing: much longer periods could 
fit the data equally well.
In that case one would have to assume that the fluxes
are higher outside the time-range covered.

The 20~cm data cannot be fitted with a 20-yr period sine function.
Either a much longer period would be needed, or the 
data do not follow a sine curve. From other colliding-wind binaries,
such as WR 140 (Dougherty et al.~\cite{Dougherty+al05}), 
we know that the flux variation 
at larger wavelengths does indeed show a sharper peak than at smaller
wavelengths
(see Sect.~\ref{section colliding-wind region inside the binary}).
The behaviour of the HD~167971 20~cm fluxes seems consistent with
this and therefore does not contradict a 20-yr period.

 The 2 and 3.6~cm fluxes show some variability, but the range is small
  compared to the 6 and 20~cm range. The time coverage at 2 and
  3.6~cm is not as good as for 6~cm, but is still good enough to suggest
  that this smaller variability is real and is not due to the limited sampling.

  The spectral index between 2 (or 3.6) and 6~cm is always negative,
  and remains between $-0.3$ and $-0.6$ most of the time. The
  6--20~cm index is quite variable: 
  it is $-0.3$ around 1985, is positive
  around 1990 (if we rely on the sine interpolation of the 6~cm data),
  is $\sim$\,0.0 in 1993--1994 and is $-0.3$ again in $\sim$\,2002--2003.
  At certain times during the cycle (e.g. 1990--1994), the fluxes therefore
  no longer follow a power law, but show a turnover 
  (from negative to positive spectral index) between 6 and 20~cm.

\begin{figure}
\resizebox{\hsize}{!}{\includegraphics{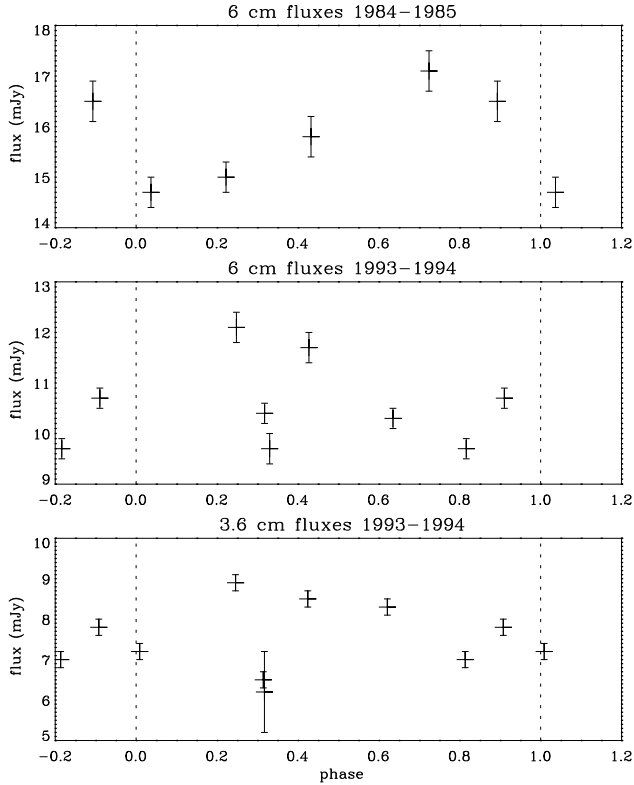}}
\caption{Various subsets of radio observations are plotted as a function
  of orbital phase, with a period of $P=3.3213$ days.
}
\label{fig 3.3 day period}
\end{figure}

\subsection{Short-term variability}
\label{section short-term variability}

Superimposed on the long-term variability, there is also
shorter-term variability with a smaller amplitude.
Clear examples of this are seen in the 6~cm 1984--1985 observations
(projects AA28, AA29 and AC116) and the 3.6 and 6~cm 1993--1994 observations 
(projects AB671 and AL320). We will consider these two intervals
separately: this has the advantage that we do not need to de-trend our
data for the long-term variations. An important question that will be
addressed is whether we can detect the 3.3-day period of the
eclipsing binary in these data.

We first of all note that these short-term
variations are significant. This is most
easily seen by determining the $\chi^2$ from fitting a constant to 
the observations. We find a reduced $\chi^2$ of 10.8 (1984--1985),
19.6 (1993--1994, 3.6~cm) and 14.0 (1993--1994, 6~cm).
 The error bars used in this calculation are from 
  Table~\ref{table radio data}.
  They cover not only the root-mean-square
  (RMS) noise in the map, but also include the uncertainty in the absolute flux
  calibration and an estimate of some of the systematic errors
  (see Paper~I). For observations with the smallest error bars,
  it is the absolute flux
  calibration that dominates.

 Fig.~\ref{fig 3.3 day period} shows that the 1984--1985 6~cm
  data show good orbital phase locking, suggesting that they
  indeed follow the 3.3-day period. However, with only 5 observations 
  this good fit may not be very significant. A series of 100 Monte 
  Carlo simulations in which the same fluxes were attributed random 
  phases showed that about 30 \% of those showed good phase locking 
  (as judged by eye). 

 For the 1993--1994 observations, neither the 3.6 nor 6~cm fluxes
  show good phase locking. Especially around phase 0.2--0.4 there
  is a large variation in flux, which is not compatible with the
  expected simple behaviour of the fluxes with orbital phase.
  If one would presume some of the observations around
  phase 0.2--0.4 to be in error and eliminate them, a further
  problem would arise: 
  the phase of minimum flux would then shift considerably
  (by $\sim 0.2$)
  between the 1984--1985 and the (remaining) 1993--1994 observations. 
To verify if such a phase shift is possible, we checked the 
optical data from the Long-term Photometry of Variables
campaign (LTPV -- Manfroid et al.~\cite{Manfroid+al91}, 
Sterken et al.~\cite{Sterken+al93}) and from Hipparcos 
(ESA~\cite{ESA97}). These data cover the time-range 1985--1993 and
show a period that is slightly different from the 
Leitherer et al.~(\cite{Leitherer+al87}) one: we find 3.32157 days
instead of 3.3213. This new value is closer to more recent determinations
of the photometric period
(3.321609~d, Mayer et al.~\cite{Mayer+al92};
3.321634~d, Van Leeuwen \& Van Genderen~\cite{VanLeeuwen+VanGenderen97}).
Using even the most
different periods (3.321634~d vs. 3.3213~d) results in a
phase shift of only 0.1 over 9 years. 
This is too small to explain the required
phase shift between the 1984--1985 and 1993--1994 data.

In summary, we conclude that the 3.3-day period of the eclipsing binary
is not detected in the radio data.

 We also searched for other periods in the data. We de-trended the 3.6 
  and 6~cm observations by subtracting the best-fit sine curve with a 20-yr 
  period. We then systematically tried periods between 3.3~d and 20~yr
  and evaluated the goodness of phase locking with the minimum
  string-length method (Dworetsky~\cite{Dworetsky83}). The most promising
  solutions were evaluated by eye, but no period was found that gave
  good phase locking for both the 3.6 and 6~cm data (in all cases, at
  least one data point was significantly discrepant).

\begin{table}
\centering
\caption{The top panel (a) shows typical parameters for
an O5~V, an O8~V and an O8~I star, based on 
stars of similar spectral type found in Puls et al.~(\cite{Puls+al96}), 
Repolust et al.~(\cite{Repolust+al04}) and
Markova et al.~(\cite{Markova+al04}).
We list the radius ($R_*$), mass ($M_*$), terminal velocity ($v_\infty$) and
mass-loss rate ($\dot{M}$). 
The bottom panel (b) shows the orbital parameters, both for the
inner binary and for the third component assumed to be
orbiting the inner binary.
In both cases, we list two possibilities for the spectral types of
the inner binary components.
We list the semi-major axis 
($a_{1,2}$) and the amplitude of the orbital motion ($K_{1,2}$),
which were calculated from the mass listed in the top panel
and an assumed inclination
angle of 90\degr. 
  The masses of an O8~V + O8~V binary are rather low;
  they are not compatible with the order of $100$~M$_{\sun}$
  for the total mass of the binary
  derived by Leitherer et al.~(\cite{Leitherer+al87}) from the observed
  velocity amplitude of $\sim$\,300~km s$^{-1}$.
For the inner binary the known period 
was used, while for the inner binary + O8~I system, 
a 20-yr period was assumed. 
}
\label{table typical parameters}
\begin{tabular}{lrrrrrrrrrrrrr}
\multicolumn{5}{c}{(a) Star and wind parameters for each component} \\
\hline\hline
\multicolumn{1}{c}{spectral} & \multicolumn{1}{c}{$R_*$} & 
   \multicolumn{1}{c}{$M_*$} & \multicolumn{1}{c}{$v_\infty$} & 
   \multicolumn{1}{c}{\rule[0mm]{0mm}{4mm}$\dot{M}$} \\
\multicolumn{1}{l}{type} & \multicolumn{1}{c}{(R$_{\sun}$)} & 
   \multicolumn{1}{c}{(M$_{\sun}$)} & \multicolumn{1}{c}{(km s$^{-1}$)} & 
   \multicolumn{1}{c}{($10^{-6}$ M$_{\sun}$ yr$^{-1}$)} \\
\hline
O5 V & 14 & 45 & 3000 & 1.5 \\
O8 V & ~9 & 13 & 2000 & 0.1 \\
\hline
O8 I & 23 & 50 & 2500 & 5.0 & \\
\hline
\end{tabular}\vspace{3mm}

\begin{tabular}{llrrrrrrrrrrrr}
\multicolumn{6}{c}{(b) Orbital parameters} \\
\hline\hline
& \multicolumn{1}{l}{spectral type} & 
    \multicolumn{1}{c}{$a_1$} & \multicolumn{1}{c}{$K_1$} &
    \multicolumn{1}{c}{$a_2$} & \multicolumn{1}{c}{$K_2$} \\
& & 
  \multicolumn{1}{c}{(R$_{\sun}$)} & \multicolumn{1}{c}{(km s$^{-1}$)} &
  \multicolumn{1}{c}{(R$_{\sun}$)} & \multicolumn{1}{c}{(km s$^{-1}$)} \\
\hline
\multicolumn{3}{l}{Inner binary}\\
& O5 V + O5 V & 21   & 320 & 21 & 320 \\
& O8 V + O8 V & 14   & 210 & 14 & 210 \\
\hline
\multicolumn{2}{l}{Triple system} & \multicolumn{2}{c}{binary} & \multicolumn{2}{c}{third component} \\
& O5 V + O5 V + O8 I & 2900 & 20  & 5300 & 37 \\
& O8 V + O8 V + O8 I & 4400 & 31  & 2300 & 16 \\
\hline
\end{tabular}

\end{table}

\section{Interpretation}
\label{section interpretation}

\subsection{Colliding-wind region inside the binary}
\label{section colliding-wind region inside the binary}

Here, we consider the possible colliding-wind region inside the binary.
Because both binary components are of equal spectral type 
(Leitherer et al.~\cite{Leitherer+al87}), the collision will occur
at the mid-plane between the two stars,
i.e. at 21~$R_{\sun}$ for an O5~V + O5~V binary (see
Table~\ref{table typical parameters}a).
If this colliding-wind region has an important contribution to the total flux, 
the 3.3-day binary period may be detectable in the fluxes. 

We did not find the 3.3-day period in the observations, however.
Two possible explanations can be suggested for this: either 
there is little or no intrinsic synchrotron emission,
or there is too much free-free absorption for the synchrotron 
emission to escape.
A lack of intrinsic emission could have various causes:
the colliding-wind region could be too weak to
provide enough energy for the synchrotron emission, the Fermi acceleration
mechanism could be inefficient, or the inverse-Compton cooling could
be too strong. We explore each of these possibilities in detail below.

  Even for such a close binary, the velocities of the material between
  both components are supersonic and
  a colliding-wind region is formed (as 
  X-ray observations of similar systems, such as the 3.4-day period 
  \object{HD~159176} confirm;
  De Becker et al.~\cite{DeBecker+al04}).
  Radiative inhibition, however, 
  considerably reduces the outflow velocities.
  Stevens \& Pollock~(\cite{Stevens+Pollock94}) calculated the effect
  of radiative inhibition for a somewhat similar binary
  (\object{HD 160652}, O6.5~V + O6.5~V, $P$ = 6.14~d). Extrapolating
  the results of their Fig.~4 to our even closer system, we estimate that
  the outflow velocity at the shock is $v \approx 500$~km~s$^{-1}$.

We first check that the collision between the winds is energetic enough
to provide the synchrotron emission. 
In this order-of-magnitude analysis, we assume HD~167971 to be
  an O5~V + O5~V binary with star and wind parameters 
  based on literature values for similar stars
  (see Table~\ref{table typical parameters}). 
The total kinetic energy of both winds is $\dot{M} v^2 = 2.4 \times 10^{35}$ erg s$^{-1}$.
Assuming the size of the emission region is $\sim$\,$\pi r_\mathrm{OB}$ (Eichler \& Usov~\cite{Eichler+Usov93}),
where $r_\mathrm{OB} = 21 R_{\sun}$, the fractional solid angle
of the stellar winds
impinging on the non-thermal emission region is $\sim0.23$, and the
kinetic luminosity available to the wind-collision region is
$5.5 \times 10^{34}$ erg s$^{-1}$. We estimate a total radio luminosity of
$\sim 1.6 \times 10^{30}$ erg s$^{-1}$ at a distance of 2 kpc, which can easily be
provided from the kinetic luminosity of the winds.

For the Fermi acceleration to be efficient, the Alfv\'en Mach number 
\begin{equation}
M_\mathrm{A} = \frac{v \sqrt{4 \pi \rho}}{B}
\end{equation}
should not be too small 
(Wentzel~\cite{Wentzel74}; 
Jones \& Ellison~\cite{Jones&Ellison91}, Sect.~4.1.1).
$B$ is the magnetic field at the shock and 
$v$ and $\rho$ are the pre-shock velocity and density.
Assuming a value of $M_\mathrm{A} = 4$
and a pre-shock density of
$7.0 \times 10^{-14}$ g cm$^{-3}$,
we find that B should be 12~G. Assuming a Weber \& Davis~(\cite{Weber+Davis67}) type magnetic field structure, this
leads to a surface field that is small compared to the usual
assumption of a 100~G surface field (compatible with the 
detection limit, see Mathys~\cite{Mathys99}).

We next explore the possibility that inverse-Compton (IC) cooling is 
too strong because of the large supply of UV photons available in this
close binary. These
photons IC scatter off the relativistic electrons and thereby 
remove energy.
The time needed for IC cooling to reduce 
the momentum of an electron 
from a high value $p_\mathrm{c}$ down to $p$ can be found by integrating
Eq.~(6) of Van Loo et al.~(\cite{VanLoo+al05}):
\begin{equation}
\Delta t = \frac{3 \pi m_\mathrm{e}^2 c^3 r^2}{p \sigma_\mathrm{T} L_*} ,
\end{equation}
where we assumed that $p \ll p_\mathrm{c}$.
As we typically consider
radio fluxes in the 2 -- 20 cm wavelength range, we
determine
the time required to cool down to a momentum that has maximum
synchrotron emission just beyond 20~cm. 
Using the relation between momentum and the frequency at which the maximum
flux is emitted (Van Loo et al., their Eq.~(19)),
we find $\Delta t \approx 600$~s (for an assumed $L_*=10^6 L_{\sun}$).
During that time, the electron has moved over a distance 
$v \Delta t \approx 0.5$~$R_{\sun}$.

Due to the IC cooling, the
size of the synchrotron emitting region is therefore rather small.
This
makes it doubtful whether the colliding-wind
region inside the binary can generate a significant amount of synchrotron
emission.
Redoing the analysis for later-type components in
the binary, or (super)giants, results in a similar 
size for the synchrotron emitting region (within a factor of three).
If the components are O5 supergiants, then we note that
the Alfv\'en number requirement
results in a $\sim 40$~G magnetic field at the shock, which translates into
a surface field that is in better agreement
with an assumed surface field of 100~G.

\begin{figure}
\resizebox{\hsize}{!}{\includegraphics{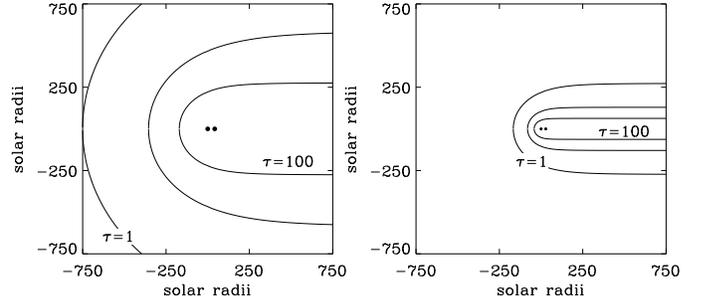}}
\caption{Contours of optical depth at 6~cm
in the stellar wind of an O5~V star (left panel)
and an O8~V star (right panel). The observer is situated to the left.
The contour lines of $\tau$ = 1, 10 and 100 are shown. The two dots
indicate the position of both components of the binary.}
\label{fig optical depth}
\end{figure}

We finally explore the possibility
that the free-free
absorption is so large that all of the synchrotron emission is
absorbed at all phases in the orbit. 
In Fig.~\ref{fig optical depth}, we show the contours of optical depth
($\tau$) at 6~cm. To simplify the figure and discussion, we consider the
free-free effects of only
a single star. 
The optical depth was calculated from
the formulae given in Wright \& Barlow~(\cite{Wright+Barlow75}).
Fig.~\ref{fig optical depth} also shows the position of both components.
For the purposes of this figure we have assumed that the inclination
is exactly 90\degr~(we know it is close to 90\degr, because
it is an eclipsing binary).
The size of the colliding-wind region is somewhat larger
than the binary separation 
and all of the synchrotron radiation will be found close to this
region (see above).
We see on Fig.~\ref{fig optical depth} that
the synchrotron-emitting region is well within
the $\tau=10$ contour surface and would therefore be 
undetectable\footnote{
  One should be careful with this conclusion. In the WR~140 system,
  for instance, the colliding-wind region is well within the free-free radius
  of the Wolf-Rayet star during the whole orbit. The O star wind blows a 
  low-density region in the Wolf-Rayet wind and it is through this region
  that we can see the synchrotron emission, at least part of the time
  (Williams et al.~\cite{Williams+al90}). The winds of the HD~167971 binary
  are nearly equal, and therefore no such complications arise here.
}.
(If the components are both giants or supergiants
  (Davidge \& Forbes~\cite{Davidge+Forbes88}) rather than 
  main-sequence stars, then the opacity
  effects discussed here will be even stronger.)

Another possibility is the intrinsic synchrotron emission from the
wind-collision region in the binary is absorbed by free-free opacity of the
O8 supergiant wind. However, as the relative positions of the supergiant
and the binary system are unknown this remains an open question.

\subsection{Colliding-wind region between the supergiant and the binary}

 The major part of the observed non-thermal emission of HD~167971
  is not due
  to the colliding-wind region inside the binary, because
  (i) we do not detect the 3.3-day period, 
  (ii) the intrinsic synchrotron flux is estimated to be small, and 
  (iii) the free-free absorption is large. 
  We can therefore
conclude that
the main source of the observed non-thermal radio emission must be the
colliding-wind region between the O8 supergiant and the binary.
We recall that from spectroscopic observations 
(Leitherer et al.~\cite{Leitherer+al87}) 
it is not clear if the supergiant and
the binary are gravitationally bound. The supergiant could be coincidentally
in the same line of sight and might even be at a (somewhat) different 
distance. However, we will assume in our analysis that the supergiant and the 
eclipsing
binary do form a gravitationally bound system. This allows us to attribute
the long-term variability to the 
relative motion of the binary and the supergiant. 

The orbital period is then $\sim$\,20 years, or possibly longer
(Sect.~\ref{section long-term variability}).
With the typical stellar parameters for an O8 supergiant and the
eclipsing binary, a 20-year period corresponds to a 
  separation 
  $a_1 + a_2 \ge 6700~R_{\sun}$ 
  (see Table~\ref{table typical parameters}b),
  provided we assume a nearly-circular orbit. 
This is well beyond the $\sim$\,1600~R$_{\sun}$
radius at which $\tau=1$ for 6~cm
in the supergiant wind.
Most of the time we therefore see the intrinsic synchrotron emission
at this wavelength. 

At 20~cm, the $\tau=1$ radius is much larger 
($\sim$\,3500~R$_{\sun}$). 
A much longer part of the orbit will therefore be spent inside the
20~cm radio photosphere (except for inclinations near 0\degr), 
explaining why the maximum at 20~cm is much
more peaked. At wavelengths shorter than 6~cm,
the radio photosphere is much smaller
and the colliding-wind region is therefore outside the radio
photosphere nearly all the time, explaining the lack of substantial
variation at 2 and 3.6~cm.
  The observed changes in the 6--20~cm spectral index 
  and the turnover between 6 and 20~cm during part of the period
  (Sect.~\ref{section long-term variability}) can thus be ascribed to these
  opacity
  effects. An alternative explanation for the turnover would be
  the Razin effect. Due to this effect, the synchrotron emission deviates
  from a power law at longer wavelengths and the long-wavelength spectral index
  can become zero or even positive. The strength of the Razin effect depends
  on the (local) magnetic field and electron density. In an eccentric orbit,
  both quantities would vary with orbital phase, and therefore, so would
  the spectral index. More sophisticated models would be required to
  derive quantitative orbital information from this alternative
  explanation of the turnover.

Eichler \& Usov~(\cite{Eichler+Usov93}) made a simple model for the
synchrotron radiation emitted in a colliding-wind binary.
When we apply their analysis to the present data,
we find an intrinsic synchrotron flux (i.e. without free-free absorption)
of $\sim$\,25~mJy at 6~cm, which is
in surprisingly good agreement with the observed values.
     The Eichler \& Usov model does contain a number
     of unconstrained parameters for which we adopted their default
     values. Hence, the agreement should only be interpreted as
     showing the plausibility of our model.

We next try to explain the variability 
  of the flux and the turnover between 6 and 20~cm (which occurs during
  part of the orbit) by ascribing them
to the changing free-free absorption due to the orbital motion of the
synchrotron-emitting source and the O8 supergiant.
We used a simple 
model, where we assumed the intrinsic synchrotron
emission to come from a point source in a 20-yr circular orbit around
the supergiant
(with stellar parameters as in Table~\ref{table typical parameters}a). 
Simulations show that we can reproduce the 6~cm
radio lightcurve quite well, provided we use a low-inclination orbit
(otherwise the flux would be completely absorbed).
However, free-free absorption increases considerably from 6 to 20~cm,
and we cannot match the 6 and 20~cm data with the same model.
This is most probably due to our assumption that the synchrotron-emitting
region is a point source. In reality, the 
synchrotron-emitting
region will 
be extended and deriving
  quantitative information about the opacity from the
  flux or the turnover will therefore require more
  sophisticated models, which are beyond the scope of this paper.

Further information on the size of the 
synchrotron-emitting region comes from the
Phillips \& Titus~(\cite{Phillips+Titus90}) observations. They
found a size of 16 mas, which corresponds to
a diameter of $\sim$\,6500~R$_{\sun}$
at 18~cm. 
 This is comparable to the separation between the binary and the
O supergiant.
The size of the colliding-wind region makes it difficult to put constraints
on the orbital inclination. 
  If the inclination of the orbit is close to 90\degr,
  a point-like source of 
  synchrotron emission would be eclipsed at 6~cm during $\sim$\,2~years 
  of the 20-year orbit,
  as it passes behind the free-free radius of the supergiant; but this
  is contradicted by the observations.
However, for a larger-sized colliding-wind region,
this argument cannot be used, as there is probably always some part of it that
is not eclipsed. No constraints on the orbital inclination can therefore be
derived.

The shorter-term variability could be attributed to stochastic
variations due to clumping in the stellar winds, as suggested for
WR 146 and WR 147 
(Setia Gunawan et al.~\cite{SetiaGunawan+al00}, \cite{SetiaGunawan+al01}).
These affect the synchrotron emitting 
region itself (as the density and/or velocity
of the material entering the colliding-wind region changes)
and affect the free-free absorption.
The AB671 series represents a typical example of such flux variations.
The 6~cm fluxes range between 10~mJy and 12~mJy. We can derive an estimate
of how much the density needs to change to explain this flux variation
through additional free-free absorption. If, for simplicity, we assume the
synchrotron emitting region to be a point source, we can use the 
Wright \& Barlow~(\cite{Wright+Barlow75}) formulae to determine how much
optical depth has to be added to change the 12~mJy flux into 10~mJy. When
the free-free optical depth along the line-of-sight is 1, a 10 \% increase
in density suffices. For larger-sized sources, the density contrasts will
need to be considerably larger to offset the cancellations that occur
in a stochastically clumped wind. 

The model proposed here is also compatible with the X-ray data,
studied by De Becker et al.~(\cite{DeBecker+al05b}).
From {\em XMM-Newton} observations,
     they found an excess of a factor of
     4, compared to the canonical $L_{\rm X}$ vs.
     $L_{\rm bol}$ relation for O stars.
     The emission
is probably thermal
and shows the presence of a high-temperature plasma component
($\sim$\,20~MK). 
The colliding-wind region within the eclipsing
binary is unlikely to produce such high plasma temperatures as the winds
of the two O5--O8~V stars collide before they reach their terminal
velocities, thereby producing softer thermal X-ray emission. The
high plasma temperature should therefore be ascribed to material heated
in the colliding-wind region between the eclipsing binary and the 
O8 supergiant.
 The existence of this interaction shows that the supergiant must be
 at approximately the same distance as the binary, and is not an
 accidental line-of-sight object.

The two ROSAT data points
plotted on Fig.~5 of De Becker et al.~(\cite{DeBecker+al05b})
show 
X-ray flux decreasing between 1993 and 1995. This 
     decline closely follows the radio emission during the same period. 
     We attribute this to the intrinsic X-ray and synchrotron emission varying
in a correlated way, as they both depend on the shock strength and the 
absorption in the stellar wind material.
However, De Becker et al. caution that the observed
variability might be attributed to
systematic effects between {\em XMM-Newton} and ROSAT data,
as seen in Cyg~OB2 No.~8A
(De Becker et al.~\cite{DeBecker+al05a}).

\section{Conclusions}
\label{section conclusions}

In this paper we analysed the available radio data on 
HD~167971 from the VLA and ATCA. 
The negative spectral index confirms the
  non-thermal nature of the radio emission.
  The fluxes are also very high (compared to thermal) and they are variable.

At the core of the HD~167971 
triple system is a 3.3-day period eclipsing binary.
No modulation of the radio fluxes with this period was detected.
We can explain this by a rather low
  intrinsic synchrotron flux
  (due to a low acceleration efficiency and/or strong
       inverse-Compton cooling),
  or by
  the large amount of free-free absorption in the stellar wind material.
We therefore conclude that most of the non-thermal emission
must come from the colliding-wind region 
between the O8 supergiant and the combined winds of the binary. 

The hypothesis that the O8 supergiant and the binary 
have a colliding-wind interaction
provides a plausible explanation for many of the characteristics 
of the radio and X-ray observations.
If, furthermore, the system is gravitationally bound, the observed 
  variability can be ascribed to the orbital motion.
The proposed orbital period of $\sim$\,20 years
is large enough to be well beyond the
$\tau = 1$ free-free radius. Some absorption still occurs, explaining
the long-term variability of the radio fluxes. The high X-ray temperature
plasma component found by De Becker et al.~(\cite{DeBecker+al05b}) can also 
be ascribed to this colliding-wind region.
The problem of fitting the 6 and 20~cm radio fluxes simultaneously with
a point-source model, 
and the Phillips \& Titus~(\cite{Phillips+Titus90}) 
resolved observations suggest a colliding-wind region of substantial size.

While this model provides a good explanation for the observations,
independent verification still needs to be sought. 
Verification of our HD~167971 model can most easily be obtained from
high-resolution, high S/N optical spectra. These should
bring definite evidence that the triple system is indeed gravitationally bound
and should allow us 
to obtain better stellar and orbital parameters for the binary.
Once these data are available, quantitative models for the X-ray and
non-thermal radio emission can be developed.

\begin{acknowledgements}
 We are grateful to the referee, Sean Dougherty, for his helpful and
  constructive comments.
  We also thank Joan Vandekerckhove for his help with the reduction of the
  VLA data and
  the original observers of the data we used from the VLA archive.
We are grateful to Jan Cuypers, Herman Hensberge, Julian Pittard
and Gregor Rauw for useful discussions.
This research has made use of the SIMBAD database, operated at CDS,
Strasbourg, France and NASA's Astrophysics Data System Abstract Service.
Part of this research was carried out in the framework of the
project IUAP P5/36 financed by the Belgian Science Policy Office.
\end{acknowledgements}

\appendix

\section{Data reduction}
\label{appendix data reduction}

The data reduction was done using the Astronomical Image Processing
System (AIPS), developed by the NRAO.
The different reduction steps consist of assigning the standard fluxes
to the flux calibrators, calibration, imaging and cleaning.
These steps have been described in detail in the Appendix to Paper~I.
On some observations made with low spatial resolution,
Galactic background structure can be seen. 
This background can easily be filtered out by using the properties of 
  the Fourier transform that is an essential part of radio interferometry:
  we simply drop those visibilities that were measured on the shortest 
  baselines and thereby eliminate the large-scale background structure.
  In doing so, we took care not to remove more data than strictly needed.
Technical information on the reduction
is listed in Table~\ref{table radio data}.

In a considerable number of cases, the distance of the phase calibrator
to HD~167971 is large (9--11\degr, see Table~\ref{table radio data}).
The antenna gains could therefore be significantly different between the 
phase calibrator and target scans. 
We therefore systematically applied selfcalibration to those observations 
that have detections.
We did one round of phase-only selfcalibration (a second round does not 
make any significant difference). 
For the time-interval over which to integrate, we took the shortest 
time-interval that gave smooth phase solutions (phase scatter of less than 
$\sim$\,20\degr): this is usually 3 or 5 min.
Observations with phase scatter significantly larger 
than 20\degr~have been listed 
in the ``notes" column
to Table~\ref{table radio data}. No selfcalibration was applied to the 
AC308 observations (because of their short on-target time) and the ATCA data.

On the selfcalibrated images, the fluxes were measured by fitting
an elliptical Gaussian to the source. We checked that the resulting
fluxes are in good agreement with the maximum intensity (this is consistent
with HD~167971 being a point source on all our images).
The fluxes are
listed in Table~\ref{table radio data}. The necessity for using selfcalibration
is clearly shown by a number of observations that have substantially larger
fluxes for HD~167971 on the selfcalibrated images than on the 
non-selfcalibrated
images. The largest effect is seen in the AB671 (1993-02-01) 3.6~cm observation,
where the selfcalibrated flux is $6.2\pm 1$~mJy, while the non-selfcalibrated 
one is only 2.6~mJy.

The error bars listed in this table cover not only the root-mean-square 
(RMS) noise in the map, but also include the uncertainty in the absolute flux 
calibration 
(2 \% for the 3.6, 6, 20 and 90~cm observations and 5 \% for the
  other wavelengths)
and an estimate of some of the {\em systematic} errors 
(see Paper~I). 
For those observations with relatively small error bars, it is the
  uncertainty in the absolute flux calibration that dominates.
In those cases where HD~167971 is off-centre,
its flux was corrected for 
primary beam attenuation and beamwidth smearing.

For the observations where
HD~167971 was not detected, we assign an upper limit of 3 times the RMS,
where the RMS is measured in a small box around the source.
If the source is off-centre, this upper limit is corrected for the
primary beam attenuation and beamwidth smearing.
  A number of 20~cm observations containing HD~167971 are of such 
  low quality that they do not even provide a significant upper limit:
  BAUD (1980-07-12 and 1982-02-28), FIX (1982-05-22)
  and AT143 (1993-06-11). A number of 90~cm observations were also rejected
  for the same reason.

In some 20~cm observations HD 167971 is within the primary beam
on two images centred on different
targets. 
In that case we only list the one closest to HD 167971 because
this gives the smallest error bar. 
In most cases the flux of the
other image is compatible with the flux listed. The exceptions are:
AC116 (1984-DEC-21), where the off-centre measurement of HD~167971
is a factor of two lower than the on-centre
measurement and BP1, where the off-centre value is a factor of two higher
than the on-centre value. A smaller effect is seen for VP51, where
the off-centre value is 30 \% higher than the on-centre value.
Simulations show that the effect in the high-resolution AC116 observation
is due to noise in the map. For the low-resolution BP1
and VP51, we checked that decreasing the background even further (by throwing
away more short baselines) leads to a much better agreement.
Some caution must therefore be used in off-centre fluxes, such as the
low-resolution AC308 observations.
(Note that the values listed for AC116, BP1 and VP51 in 
Table~\ref{table radio data}
do not suffer from this problem because they
are based on the on-centre measurements.)

A number of observations have used two flux calibrators. We checked that
using either calibrator results in the same flux (within the error bars).
This gave good agreement in most cases, except AL320 (1994-FEB-18) 3.6~cm,
where there is a 32 \% difference. Checking other observations
made around the same time shows that the (uncalibrated) fluxes of \object{3C48}
are anomalously low for the observation we are interested in, while the
(uncalibrated) \object{3C286} fluxes remain the same over all the observations
considered. The calibration of this observation is therefore made on
3C286. A similar problem exists for AL320 (1994-FEB-18) 6~cm. Applying the same
technique shows that, in this case, 3C48 is the more reliable calibrator.
We also note that for the TSTOB observation we used the flux calibrator
from another programme, nearby in time (see Paper~I).

The reduction of the ATCA observations C978 was detailed in Paper~I.
These observations are centred on HD~168112, but the 
13 and 20~cm primary beams are large enough that they also contain 
HD~167971. Because these data were collected in a number of spectral
channels, we can apply multi-frequency synthesis: this results in
images that are not beamwidth smeared.

  A comparison between our fluxes and those that have been published in
  the literature shows that our error bars are
  usually larger, because we include the flux calibration errors and
  some estimate of systematic effects. 
  Our 2~cm determinations are systematically higher than the literature values.
  This is because of the improvement due to the selfcalibration 
  reduction technique.
  A similar, but smaller, effect is seen at 6~cm.
  For the 20~cm observations of Bieging et al.~(\cite{Bieging+al89}) we obtain
  lower fluxes. The VLA Calibrator Manual instructions
  (Perley \& Taylor~\cite{Perley+Taylor03})
  for this specific combination of wavelength, flux calibrator and
  VLA configuration show that the results obtained should be reduced by
  6~\%. We speculate that Bieging et al. did not apply this 6~\% reduction;
  doing so results in an acceptable agreement between their fluxes and ours.
  Our upper limit to the AR328 0.7~cm observation (4.5~mJy)
  is much higher than the Contreras et al. (\cite{Contreras+al96})
  value of 1.72~mJy, We note that this is a
  19 min observation using
  only 10 antennas, which gives a theoretical RMS of 0.7~mJy/beam. The 3 sigma
  upper limit of 1.72~mJy claimed by Contreras et al. is therefore
  too optimistic, and we have greater confidence in our result.

\small

\begin{longtable}{llllr@{ $\pm$ }lrrrllr@{ $\pm$ }ll}
\caption{\label{table radio data}Reduction of VLA and ATCA data. 
Programme C978 is an ATCA observation, all others are VLA observations.
Column (1) gives the programme name, (2) the date of the observation,
(3) the phase calibrator (J2000 coordinates),
(4) the flux of the phase calibrator $\pm$ the rms (in Jy),
(5) the distance of HD~167971 to the phase calibrator (degrees),
(6) the integration time (in minutes) on the source,
(7) the number of antennas that gave a usable signal,
(8) the configuration the VLA was in at the time of the observation,
(9) the RMS in the centre of the image (not listed in case on an upper limit),
(10) the measured flux (in mJy) and
(11) refers to the notes.
Many of the VLA observations were made in two sidebands, each of which has a 
bandwidth of 50 MHz; the exceptions are noted in column (11).
Upper limits are 3 $\times$ the RMS.
Numbers between brackets in the notes column give references for those
observations that have already been published in the literature.}\\
\hline\hline
& \multicolumn{1}{c}{(1)} & \multicolumn{1}{c}{(2)} & \multicolumn{1}{c}{(3)} &
\multicolumn{2}{c}{(4)} & \multicolumn{1}{c}{(5)} & \multicolumn{1}{c}{(6)} &
\multicolumn{1}{c}{(7)} & \multicolumn{1}{c}{(8)} & \multicolumn{1}{c}{(9)} &
\multicolumn{2}{c}{(10)} & \multicolumn{1}{c}{(11)} \\
& \multicolumn{1}{c}{progr.} & \multicolumn{1}{c}{date} &
\multicolumn{4}{c}{phase calibrator} &
\multicolumn{1}{c}{intgr.} &
\multicolumn{1}{c}{no.} &
\multicolumn{1}{c}{config} &
\multicolumn{1}{c}{RMS} &
\multicolumn{2}{c}{flux} &
\multicolumn{1}{c}{notes}\\
\cline{4-7}
&  &  &
\multicolumn{1}{c}{name} &
\multicolumn{2}{c}{flux} &
\multicolumn{1}{c}{dist.} &
\multicolumn{1}{c}{time} &
\multicolumn{1}{c}{ants.} & &
\multicolumn{1}{c}{(mJy)} &
\multicolumn{2}{c}{(mJy)} & \\
\hline
\endfirsthead
\caption{continued.}\\
\hline\hline
& \multicolumn{1}{c}{(1)} & \multicolumn{1}{c}{(2)} & \multicolumn{1}{c}{(3)} &
\multicolumn{2}{c}{(4)} & \multicolumn{1}{c}{(5)} & \multicolumn{1}{c}{(6)} &
\multicolumn{1}{c}{(7)} & \multicolumn{1}{c}{(8)} & \multicolumn{1}{c}{(9)} &
\multicolumn{2}{c}{(10)} & \multicolumn{1}{c}{(11)} \\
& \multicolumn{1}{c}{progr.} & \multicolumn{1}{c}{date} &
\multicolumn{4}{c}{phase calibrator} &
\multicolumn{1}{c}{intgr.} &
\multicolumn{1}{c}{no.} &
\multicolumn{1}{c}{config} &
\multicolumn{1}{c}{RMS} &
\multicolumn{2}{c}{flux} &
\multicolumn{1}{c}{notes}\\
\cline{4-7}
&  &  &
\multicolumn{1}{c}{name} &
\multicolumn{2}{c}{flux} &
\multicolumn{1}{c}{dist.} &
\multicolumn{1}{c}{time} &
\multicolumn{1}{c}{ants.} & &
\multicolumn{1}{c}{(mJy)} &
\multicolumn{2}{c}{(mJy)} & \\
\hline
\endhead
\hline
\endfoot
\multicolumn{14}{l}{
 \begin{tabular}{lllllllllllll}
 \multicolumn{2}{l}{Notes:} \\
 B   & Observation is not centred on HD~167971. The flux has been corrected for 
       that. \\
     & No correction for beamwidth smearing needs to be applied to C978. \\
 C   & for VP51, no flux calibrator was available, so we calibrated
       the flux on the phase calibrator, \\
     & to which we assigned the value listed 
       in Perley \& Taylor~(\cite{Perley+Taylor03}).\\
 I   & Interference is high in the AB1065 18~cm image. \\
 PH  & The phases in the selfcalibration gain solutions show scatter that is significantly larger than 20\degr. \\
 X   & The phase calibrator is of low quality (AB671, C978), is not listed in 
       the VLA Calibrator Manual \\ 
     & (Perley \& Taylor~\cite{Perley+Taylor03};
       AW515 at 1.3~cm, BB116), or the recommended constraints on the \\
     & uv-range could not be applied (AR328 at 3.6~cm). \\
 (1) & Bieging et al.~(\cite{Bieging+al89}). \\
 (2) & Phillips \& Titus~(\cite{Phillips+Titus90}). \\
 (3) & Contreras et al.~(\cite{Contreras+al96}). \\
 \end{tabular}
}
\endlastfoot
\multicolumn{3}{l}{\bf 0.7 cm}\\*
& AR328  & 1995-04-27 & \object{1733$-$130} & 11.15 & 0.15  & 11.0 & 19 & 10 & D   &        &  \multicolumn{2}{c}{$< 4.5$} & (3) \\ 
\multicolumn{3}{l}{\bf 1.3 cm}\\*
& AW515  & 1999-06-08 & \object{1832$-$105} & 1.023 & 0.015 &  3.9 & 50 & 19 & AD  & 0.22   &  6.0 & 0.6 & PH,X \\ 
\multicolumn{3}{l}{\bf 2 cm}\\*
& AA29   & 1984-04-04 & 1733$-$130 & 5.69  & 0.09  & 11.0 & 22 & 24 & C   & 0.18   &  9.1 & 0.5 & (1) \\ 
& AC116  & 1984-11-27 & 1733$-$130 & 7.3   & 0.1   & 11.0 & 28 & 25 & A   & 0.16   & 10.1 & 0.6 & PH,(1) \\ 
& AC116  & 1984-12-21 & 1733$-$130 & 7.1   & 0.2   & 11.0 & 26 & 27 & A   & 0.18   &  9.2 & 0.6 & PH,(1) \\ 
& AC116  & 1985-02-16 & 1733$-$130 & 6.56  & 0.06  & 11.0 & 37 & 25 & A   & 0.13   &  7.4 & 0.4 & PH,(1) \\ 
& AR328  & 1995-04-27 & 1733$-$130 & 9.77  & 0.06  & 11.0 & 22 & 16 & D   & 0.25   &  6.4 & 0.5 & PH,(3) \\ 
& AW515  & 1999-06-08 & 1832$-$105 & 1.25  & 0.01  &  3.9 & 30 & 20 & AD  & 0.21   &  6.8 & 0.4 & PH    \\ 
\multicolumn{3}{l}{\bf 3.6 cm}\\*
& BP1    & 1992-05-30 & 1733$-$130 & 5.00  & 0.03  & 11.0 & 55 & 24 & CD  & 0.05   &  6.5 & 0.2 &     \\ 
& AB671  & 1993-01-21 & \object{1811$-$209} & 0.179 & 0.001 &  8.8 & 42 & 22 & A   & 0.04   & 7.2 & 0.2 & X \\ 
& AB671  & 1993-01-24 & 1811$-$209 & 0.186 & 0.003 &  8.8 & 19 & 27 & A   & 0.05   &  7.8 & 0.2 & X \\ 
& AB671  & 1993-01-29 & 1811$-$209 & 0.175 & 0.001 &  8.8 & 20 & 26 & BnA & 0.05   &  8.5 & 0.2 & X \\ 
& AB671  & 1993-02-01 & 1811$-$209 & 0.174 & 0.001 &  8.8 &  7 & 27 & BnA & 0.11   &  6.2 & 1   & X \\ 
& AB671  & 1993-02-06 & 1811$-$209 & 0.179 & 0.001 &  8.8 & 19 & 27 & BnA & 0.04   &  7.0 & 0.2 & X \\ 
& AB671  & 1993-02-14 & 1811$-$209 & 0.188 & 0.001 &  8.8 &  9 & 27 & BnA & 0.10   &  8.9 & 0.2 & X \\ 
& AL320  & 1994-02-18 & 1733$-$130 & 4.85  & 0.06  & 11.0 & 12 & 26 & AC  & 0.06   &  6.5 & 0.2 & PH \\ 
& AL320  & 1994-05-10 & 1733$-$130 & 4.69  & 0.05  & 11.0 & 10 & 24 & AB  & 0.07   &  8.3 & 0.2 &     \\ 
& AR328  & 1995-04-27 & 1733$-$130 & 6.74  & 0.02  & 11.0 & 11 & 16 & D   & 0.12   &  7.5 & 0.2 & X,(3) \\ 
& AW515  & 1999-06-08 & 1832$-$105 & 1.36  & 0.01  &  3.9 & 10 & 20 & AD  & 0.11   &  8.4 & 0.2 &     \\ 
& BB116  & 1999-12-04 & \object{1822$-$096} & 1.37  & 0.02  &  2.8 &121 & 19 & B   & 0.03   & 10.0 & 0.2 & X   \\ 
& BB116  & 2000-06-26 & 1822$-$096 & 1.340 & 0.003 &  2.8 &117 & 26 & DnC & 0.03   &  9.2 & 0.2 &     \\ 
& AB1048 & 2002-03-24 & 1832$-$105 & 1.67  & 0.04  &  3.9 &  8 & 26 & A   & 0.07   & 12.3 & 0.4 &   \\ 
& AB1065 & 2002-09-11 & 1832$-$105 & 1.29  & 0.02  &  3.9 &  7 & 26 & CnB & 0.11   &  8.9 & 0.3 &     \\ 
\multicolumn{3}{l}{\bf 6 cm}\\*
& AA28   & 1984-03-09 & 1733$-$130 & 5.74  & 0.01  & 11.0 & 19 & 26 & CnB & 0.08   & 15.8 & 0.4 & (1)     \\ 
& AA29   & 1984-04-04 & 1733$-$130 & 5.69  & 0.03  & 11.0 & 11 & 27 & C   & 0.11   & 15.0 & 0.3 & (1)     \\ 
& AC116  & 1984-11-27 & 1733$-$130 & 5.02  & 0.02  & 11.0 & 21 & 24 & A   & 0.09   & 17.1 & 0.4 & (1)   \\ 
& AC116  & 1984-12-21 & 1733$-$130 & 5.05  & 0.02  & 11.0 & 41 & 26 & A   & 0.06   & 16.5 & 0.3 & (1)\\ 
& AC116  & 1985-02-16 & 1733$-$130 & 5.312 & 0.015 & 11.0 & 19 & 24 & A   & 0.09   & 14.7 & 0.3 & (1)   \\ 
& AC216  & 1988-02-27 & 1811$-$209 & 0.303 & 0.002 &  8.8 & 15 & 25 & CnB & 0.07   & 26   & 10  & B,PH    \\
& AD219  & 1988-04-15 & \object{1804$+$010} & 1.149 & 0.004 & 13.7 & 17 & 26 & CD  & 0.04   & 34   & 6 & B,PH    \\ 
& AB671  & 1993-01-24 & 1811$-$209 & 0.324 & 0.002 &  8.8 & 19 & 27 & A   & 0.05   & 10.7 & 0.2 &        \\ 
& AB671  & 1993-01-29 & 1811$-$209 & 0.318 & 0.001 &  8.8 & 19 & 26 & BnA & 0.07   & 11.7 & 0.3 &        \\ 
& AB671  & 1993-02-01 & 1811$-$209 & 0.314 & 0.001 &  8.8 & 13 & 27 & BnA & 0.07   & 10.4 & 0.2 &        \\ 
& AB671  & 1993-02-06 & 1811$-$209 & 0.322 & 0.001 &  8.8 & 19 & 27 & BnA & 0.06   &  9.7 & 0.2 &     \\ 
& AB671  & 1993-02-14 & 1811$-$209 & 0.319 & 0.001 &  8.8 & 10 & 27 & BnA & 0.08   & 12.1 & 0.3 &        \\ 
& AL320  & 1994-02-18 & 1733$-$130 & 5.09  & 0.06  & 11.0 &  9 & 26 & AC  & 0.09   &  9.7 & 0.3 & PH     \\ 
& AL320  & 1994-05-10 & 1733$-$130 & 4.43  & 0.02  & 11.0 & 10 & 26 & AB  & 0.08   & 10.3 & 0.2 &         \\ 
& AR328  & 1995-04-27 & 1733$-$130 & 5.04  & 0.02  & 11.0 & 11 & 16 & D   & 0.20   &  8.6 & 0.4 & (3)     \\ 
& AW515  & 1999-06-08 & 1832$-$105 & 1.200 & 0.015 &  3.9 & 10 & 20 & AD  & 0.16   & 10.5 & 0.4 &         \\ 
& BB116  & 2000-06-26 & 1822$-$096 & 2.25  & 0.03  &  2.8 & 93 & 26 & DnC & 0.05   & 11.5 & 0.3 & X       \\ 
& AB1048 & 2002-03-24 & 1832$-$105 & 1.32  & 0.01  &  3.9 &  6 & 26 & A   & 0.09   & 15.1 & 0.3 &         \\ 
& AB1065 & 2002-09-11 & 1832$-$105 & 1.26  & 0.07  &  3.9 &  5 & 26 & CnB & 0.14   & 13.4 & 0.3 &         \\ 
\multicolumn{3}{l}{\bf 13 cm}\\*
& C978   & 2001-10-11 & 1832$-$105 & 1.042 & 0.002   &  3.9 &120 &  6 &EW352& 0.26  & \multicolumn{2}{c}{11 $\pm$ 2} & B,X \\
\multicolumn{3}{l}{\bf 20 cm}\\*
& AC116  & 1984-11-27 & 1733$-$130 & 5.55  & 0.03  & 11.0 & 28 & 25 & A   & 0.12  & 24.3 & 0.6 & PH,(1)           \\
& AC116  & 1984-12-21 & 1733$-$130 & 5.90  & 0.04  & 11.0 & 10 & 27 & A   & 0.17  & 24.9 & 0.6 & (1)             \\
& AG163  & 1984-12-24 & \object{1743$-$038} & 1.281 & 0.015 & 11.9 &  1 & 27 & A   &       & \multicolumn{2}{c}{$< 100 $} & B,2x25         \\
& AC116  & 1985-02-16 & 1733$-$130 & 6.36  & 0.03  & 11.0 & 15 & 25 & A   & 0.12  & 22.0 & 0.5 & (1)             \\
& AB531  & 1989-05-01 & \object{1834$-$126} & 0.518 & 0.001 &  3.9 &  2 & 26 & B   &       & \multicolumn{2}{c}{$< 100 $} & B,1x50            \\
& VP51   & 1989-06-19 & 1733$-$130 & \multicolumn{2}{c}{5.200} & 11.0 & 59 & 25 & C   & 0.24  &  6.9 & 0.3 & PH,X,C               \\
& AG290  & 1989-07-20 & \object{1833$-$210} & 13.02 & 0.03  &  9.6 &  0 & 27 & CD  &       & \multicolumn{2}{c}{$<  80 $} & B              \\
& VP91   & 1989-11-10 & 1733$-$130 & 6.2   & 0.2   & 11.0 & 56 & 27 & D   & 1.5   &  4.2  &   2 & PH,X,(2)             \\
& BP1    & 1992-05-30 & \object{1911$-$201} & 3.0   & 0.2   & 15.0 & 91 & 25 & CD  & 0.8   &  5.6 &   3 & PH                \\
& TSTOB  & 1993-07-05 & 1822$-$096 & 4.3   & 0.2   &  2.8 & 13 & 27 & C   & 0.3  &  12.9  & 0.9 &             \\
& AS534  & 1994-05-21 & 1822$-$096 & 5.39  & 0.04  &  2.8 &  4 & 21 & BnA &       & \multicolumn{2}{c}{$<  70   $} & B,1x3.125       \\
& AS534  & 1994-05-26 & 1822$-$096 & 4.99  & 0.03  &  2.8 &  5 & 25 & BnA &       & \multicolumn{2}{c}{$<  34   $} & B,X,1x3.125      \\
& AS534  & 1994-06-01 & 1822$-$096 & 5.15  & 0.02  &  2.8 &  4 & 25 & BnA &       & \multicolumn{2}{c}{$< 150   $} & B,X,1x3.125 \\
& AC308  & 1996-06-08 & 1833$-$210 & 10.78 & 0.02  &  9.6 &  1 & 27 & DnC &       & \multicolumn{2}{c}{$< 190   $} & B            \\
& AC308  & 1996-06-09 & 1833$-$210 & 10.80 & 0.02  &  9.6 &  1 & 27 & DnC & 2.0   & 12 & 4 & B                 \\
& AC308  & 1996-06-16 & 1833$-$210 & 10.78 & 0.02  &  9.6 &    & 27 & DnC &       & \multicolumn{2}{c}{$<  70   $} & B            \\
& AC308  & 1997-10-16 & 1833$-$210 & 10.07 & 0.02  &  9.6 &  1 & 27 & DnC & 2.0   & 12 & 6 & B                 \\
& C978   & 2001-10-11 & 1832$-$105 & 0.929 & 0.001 &  3.9 &120 &  6 &EW352& 0.42  & \multicolumn{2}{c}{14 $\pm$ 2} & B,X \\
& AB1048 & 2002-03-24 & 1834$-$126 & 0.534 & 0.001 &  3.9 &  3 & 26 & A   & 0.20  & 20.9 & 0.6 &               \\
& AB1065 & 2002-09-11 & 1834$-$126 & 0.468 & 0.001 &  3.9 &  5 & 26 & CnB & 6     & 29   &  23 & B,I,18cm       \\
& AB1065 & 2002-09-11 & 1834$-$126 & 0.529 & 0.001 &  3.9 &  5 & 21 & CnB & 3     & 27   &   7 & B,20cm            \\
\multicolumn{3}{l}{\bf 90 cm}\\*
& AH299  & 1988-06-21 & \object{1829$+$487} & 42.6  & 0.5   & 61.0 & 47 & 21 & DnC &       & \multicolumn{2}{c}{$< 270   $} & B,2x3.125       \\
& AH299  & 1989-05-28 & 1829$+$487 & 41.7  & 0.5   & 61.0 & 57 & 25 & CnB &       & \multicolumn{2}{c}{$<  80   $} & B,2x3.125       \\
\hline
\end{longtable}


\begin{thebibliography}{}
\bibitem[1978]{Bell78} Bell, A. R. 1978, \mnras, 182, 147
\bibitem[1989]{Bieging+al89} Bieging, J. H., Abbott, D. C., \& 
               Churchwell, E. B. 1989, ApJ, 340, 518
\bibitem[2005]{Blomme+al05} Blomme, R., Van Loo, S., De Becker, M., et al. 
               2005, A\&A, 436, 1033 (Paper~I)
\bibitem[1996]{Contreras+al96} Contreras, M. E., Rodr\'{\i}guez, L. F.,
               G\'{o}mez, Y., \& Vel\'{a}zquez, A. 1996, ApJ, 469, 329
\bibitem[1997]{Contreras+al97} Contreras, M. E., Rodr\'{\i}guez, L. F., 
               Tapia, M., et al. 1997, ApJ, 488, L153
\bibitem[1988]{Davidge+Forbes88} Davidge, T. J., Forbes, D. 1988,
              MNRAS, 235, 797
\bibitem[2004]{DeBecker+al04} De Becker, M., Rauw, G., Pittard, J. M., et al.
               2004, A\&A, 416, 221
\bibitem[2005a]{DeBecker+al05a} De Becker, M., Rauw, G., \& Swings, J.-P.
               2005a, Ap\&SS, 297, 291
\bibitem[2005b]{DeBecker+al05b} De Becker, M., Rauw, G., Blomme, R., et al.
               2005b, A\&A, 437, 1029
\bibitem[2000]{Dougherty+Williams00} Dougherty, S. M., \& Williams, P. M.
               2000, MNRAS, 319, 1005
\bibitem[2003]{Dougherty+al03} Dougherty, S. M., Pittard, J. M., Kasian, L.,
               et al. 2003, A\&A, 409, 217
\bibitem[2005]{Dougherty+al05} Dougherty, S. M., Beasley, A. J., 
               Claussen, M. J., Zauderer, B. A., \& Bolingbroke, N. J.
               2005, ApJ, 623, 447
\bibitem[1983]{Dworetsky83} Dworetsky, M. M. 1983, MNRAS, 203, 917
\bibitem[1993]{Eichler+Usov93} Eichler, D., \& Usov, V. 1993,
               ApJ, 402, 271
\bibitem[1997]{ESA97} ESA 1997, The Hipparcos and Tycho Catalogues, 
               ESA SP-1200
\bibitem[1978]{Humphreys78} Humphreys, R. M. 1978, ApJS, 38, 309
\bibitem[1991]{Jones&Ellison91} Jones, F. C., \& Ellison, D. C. 1991,
               Space Sci. Rev. 58, 259
\bibitem[1987]{Leitherer+al87} Leitherer, C., Forbes, D., Gilmore, A. C.,
              et al. 1987, A\&A, 185, 121
\bibitem[1991]{Manfroid+al91} Manfroid, J., Sterken, C., Bruch, A., et al.
               1991, A\&AS, 87, 481
\bibitem[2004]{Markova+al04} Markova, N., Puls, J., Repolust, T., \& 
               Markov, H. 2004, A\&A, 413, 693
\bibitem[1999]{Mathys99} Mathys, G. 1999, in Variable and Non-spherical
               Stellar Winds in Luminous Hot Stars, ed. B. Wolf, O. Stahl, \& A.
               W. Fullerton, Lect. Notes Phys., 523, 95
\bibitem[1992]{Mayer+al92} Mayer, P., Dreschsel, H., \& Lorenz, R. 1992,
               IBVS, 3805, 1
\bibitem[1994]{Miralles+al94} Miralles, M. P., Rodr\'{i}guez, L. F., Tapia, M.,
               et al. 1994, A\&A, 282, 547
\bibitem[1988]{Owocki+al88} Owocki, S. P., Castor, J. I., \& Rybicki, G. B.
               1988, ApJ, 335, 914
\bibitem[2003]{Perley+Taylor03} Perley, R. A., \& Taylor, G. B. 2003,
               The VLA Calibrator Manual \\
               {\tt http://www.aoc.nrao.edu/$\sim$gtaylor/calib.html}
\bibitem[1990]{Phillips+Titus90} Phillips, R. B., \& Titus, M. A.
              1990, ApJ, 359, L15
\bibitem[2006]{Pittard+al06} Pittard, J. M., Dougherty, S. M., Coker, R. F.,
              O'Connor, E., \& Bolingbroke, N. J. 2005, A\&A, 446, 1001
\bibitem[1996]{Puls+al96} Puls, J., Kudritzki, R. P., Herrero, A., et al.
              1996, A\&A, 305, 171
\bibitem[2004]{Repolust+al04} Repolust, T., Puls, J., \& Herrero, A.  2004, 
               A\&A, 415, 349
\bibitem[2000]{SetiaGunawan+al00} Setia Gunawan, D. Y. A., De Bruyn, A. G., 
              Van der Hucht, K. A., \& Williams, P. M. 2000, A\&A, 356, 676
\bibitem[2001]{SetiaGunawan+al01} Setia Gunawan, D. Y. A., De Bruyn, A. G., 
              Van der Hucht, K. A., \& Williams, P. M.  2001, A\&A, 368, 484
\bibitem[1993]{Sterken+al93} Sterken, C., Manfroid, J., Anton, K., et al.
              1993, A\&AS, 102, 79
\bibitem[1994]{Stevens+Pollock94} Stevens, I. R., \& Pollock, A. M. T.
               1994, MNRAS, 269, 226
\bibitem[1997]{VanLeeuwen+VanGenderen97} Van Leeuwen, F., \& Van Genderen, A. M.
              1997, A\&A, 327, 1070
\bibitem[2005]{VanLoo+al05} Van Loo, S., Runacres, M. C., \&
               Blomme, R. 2005, A\&A, 433, 313
\bibitem[2006]{VanLoo+al06} Van Loo, S., Runacres, M. C., \&
               Blomme, R. 2006, A\&A, 452, 1011
\bibitem[1967]{Weber+Davis67} Weber, E. J., \& Davis, L., Jr. 1967, 
               ApJ, 148, 217
\bibitem[1974]{Wentzel74} Wentzel, D. G. 1974, ARA\&A, 12, 71
\bibitem[1990]{Williams+al90} Williams, P. M., Van der Hucht, K. A., 
              Pollock, A. M. T., et al. 1990, MNRAS, 243, 662
\bibitem[1975]{Wright+Barlow75} Wright, A. E., \& Barlow, M. J. 1975,
               MNRAS, 170, 41
\end{thebibliography}
\end{document}